\documentclass[12pt,preprint]{aastex}
\usepackage{bm,epsfig,graphics,graphicx}

\shortauthors{HUGHES \& MENOU} 
\shorttitle{GOLDEN BINARIES}

\def\gsim{\;\rlap{\lower 2.5pt
 \hbox{$\sim$}}\raise 1.5pt\hbox{$>$}\;}
\def\lsim{\;\rlap{\lower 2.5pt
   \hbox{$\sim$}}\raise 1.5pt\hbox{$<$}\;}

\def\btheta{\mbox{\boldmath$\theta$}}
\def\bGamma{\mbox{\boldmath$\Gamma$}}

\begin{document}

\title{Golden Binary Gravitational-wave Sources: Robust Probes of
Strong-Field Gravity}

\author{Scott A.\ Hughes\altaffilmark{1} \& Kristen Menou\altaffilmark{2}}

\altaffiltext{1}{Department of Physics and Center for Space Research,
Massachusetts Institute of Technology, 77 Massachusetts Avenue,
Cambridge, MA 02139, USA}

\altaffiltext{2}{Department of Astronomy, Columbia University, 550
West 120th Street, New York, NY 10027, USA}


\begin{abstract}
Space-born gravitational-wave interferometers such as {\it LISA} will
detect the gravitational wave (GW) signal from the inspiral, plunge
and ringdown phases of massive black hole binary mergers at
cosmological distances.  From the inspiral waves, we will be able to
measure the masses of the binaries' members; from the ringdown waves,
we will be able to measure the mass of the final merged remnant.  A
subset of detected events allow the identification of both the
inspiral and the ringdown waveforms in a given source, and thus allow
a measurement of the total mass-energy lost to GWs over the
coalescence, $M_{\rm GW}$.  We define ``golden'' binary mergers to be
those with measurement errors likely to be small enough for a
physically useful determination of $M_{\rm GW}$.  A detailed
sensitivity study, combined with simple black hole population models,
suggests that a few golden binary mergers may be detected during a
3-year {\it LISA} mission lifetime.  Any such mass deficit measurement would
constitute a robust and valuable observational test of strong-field
relativistic gravity. An extension of this concept to include spin
measurements may allow a direct empirical test of the black hole area
theorem.
\end{abstract}

\keywords{black hole physics -- cosmology: theory -- quasars: general
-- galaxies: active, nuclei, interactions -- gravitation --
relativity}

\section{Introduction}
\label{sec:intro}

The Laser Interferometer Space Antenna ({\it LISA}), a joint ESA and
NASA enterprise, is due for launch around the year 2013.  One of its
main science objectives is the direct detection of gravitational waves
(GWs) from coalescing massive black holes (MBHs) at cosmological
distances\footnote{{See {\tt http://lisa.nasa.gov/}}}.  Although the
mission concept has not yet been entirely finalized, it is already
possible to estimate the accuracy with which {\it LISA} will measure a
variety of observables related to MBH coalescences. For example,
several groups have made predictions for the rate of MBH mergers
detectable by {\it LISA} under various assumptions for the underlying
cosmological model of BH mass assembly (see, e.g., Haehnelt 1994;
Menou, Haiman \& Narayanan 2001; Wyithe \& Loeb 2003; Islam, Taylor \&
Silk 2004; Sesana et al. 2004a,b).

These measurements will provide rich information about the binary
generating the GW signal.  Hughes (2002) has presented detailed
calculations of the exquisite precision with which {\it LISA} will
measure MBH masses in equal-mass mergers, as a function of redshift.
Vecchio (2004) has shown that Hughes' estimates are in fact fairly
pessimistic: by taking into account spin-induced precessional effects
in the binary coalescence, mass measurements can be improved.  It
should even possible to extract useful information about the spins of
the binary's members.

In this analysis, we focus on a specific class of MBH binaries that we
label ``golden'' binaries.  These are binaries for which {\it LISA}
will witness the entire coalescence, from inspiral to plunge and
ensuing ringdown.  In these events, the masses of the binary's BH
components before merger and of the final merged remnant can each be
measured to high precision.  Because GWs carry mass-energy from the
system, the system's final mass $M_f$ will be less than the initial
mass $M_i$ it has when the binary's members are widely separated.  For
golden binaries, the masses are determined so precisely that the total
mass deficit due to GW emission, $M_{\rm GW} \equiv M_i - M_f$, can
potentially be determined observationally.

Such a measurement would constitute a very robust test of a strikingly
strong-field prediction of general relativity.  At present, theory
does not tell us us too much about $M_{\rm GW}$ in general, though we
have information about some idealized cases, such as extreme mass
ratios (e.g., Davis et al 1971, Nakamura, Oohara, \& Kojima 1987).
Much effort is being directed towards improving this situation, both
via large scale numerical efforts (see Baumgarte \& Shapiro 2003 for a
review, or Br\"ugmann, Tichy, \& Jansen 2004 for a snapshot of recent
progress) and analytical calculations (Damour 2001).  It is thus
obviously of great interest to understand what direct observational
constraints can be obtained about this process as it occurs in nature.

We show that for the golden binaries {\it LISA} has the capability to
measure mass deficits in cosmic MBH mergers to high enough accuracy
that strong-field general relativity can be tested.  A particularly
noteworthy feature of this measurement is that our calculation relies
only on parameters measured during the ``inspiral'' (when the binary's
members are widely separated and slowly spiraling towards one another
due to the backreaction of GW emission) and during the ``ringdown''
(the last dynamics of the system, generated after the binary's holes
have merged and are settling down to a quiescent Kerr black hole
state).  No information about the highly dynamical final plunge and
``merger'' process is needed.  In this sense, measurement of $M_{\rm
GW}$ from golden binaries is very robust, and will thus constitute a
particularly simple but powerful probe of strong-field gravity.  We
argue that it is not unreasonable to expect {\it LISA} to perform such
measurements for a 3-year mission life-span.

In {\S\ref{sec:model}}, we describe the parameterized gravitational
waveform upon which we base our analysis.  We describe in some detail
the forms we use for the inspiral and ringdown waves, outlining how
they depend upon and thus encode the binary's masses.  It's worth
noting that the functional form of the inspiral waves that we use does
not take into account precessional effects that arise due to
spin-orbit and spin-spin interactions.  A substantial improvement in
mass determination during the inspiral can be obtained by taking this
effect into account; the spins of the binary's members can thereby be
measured as well.  This comes, however, with a substantial increase in
waveform complexity and thus computational cost (Vecchio 2004).  It is
worth emphasizing this point since it is likely that our estimates of
how accurately $M_{\rm GW}$ is measured are somewhat pessimistic due
to our reliance on this relatively crude parameterization.

In {\S\ref{sec:measure}}, we describe the calculation we perform to
estimate how accurately $M_{\rm GW}$ can be measured.  What is clear
from the modeling that has been done to date is that $M_{\rm GW}$ is a
fairly small fraction of the binary's total mass --- at most,
$10-20\%$ of the total binary mass $M$ is radiated away over the
entire coalescence.  In order for a determination of $M_{\rm GW}$ to
be meaningful, the error $\delta M_{\rm GW}$ must be substantially
smaller than $M_{\rm GW}$ itself.  We (somewhat arbitrarily) define
the regime of golden binaries to be those for which $\delta M_{\rm
GW}/M \simeq 5\%$ or less.  We construct the error $\delta M_{\rm GW}$
from errors in the mass parameters measured by the inspiral [the
``chirp mass'' and reduced mass, defined in Eq.\
(\ref{eq:insp_masses})] and the ringdown (the final mass of the
system).  We outline how this error is calculated using a standard
maximum-likelihood measurement formalism (Finn 1992).  We then perform
a large number of Monte-Carlo simulations to assess how well $M_{\rm
GW}$ can be measured as a function of source masses and redshift.  We
find that binaries with total mass $M$ in the (very rough) range ${\rm
several}\times 10^5\, M_\odot \lesssim M \lesssim {\rm a\ few}\times
10^6\,M_\odot$ have the potential to be golden out to a redshift $z
\sim 3 - 4$.

In {\S\ref{sec:rates}}, we calculate event rates for golden binary
mergers.  The specific population models we use are based on a merger
tree describing dark matter halo evolution in a concordance
$\Lambda$-CDM cosmology.  This scenario suggests that {\it LISA}
should plausibly be able to measure several golden events during its
mission lifetime.  We conclude in {\S\ref{sec:conclude}} with a
discussion of the limitations of and possible extensions to our work.

\section{Model for the coalescence waves}
\label{sec:model}

As is common, we break the binary black hole coalescence process into
three more-or-less distinct epochs --- the {\it inspiral}, {\it
merger}, and {\it ringdown}.  We briefly describe these three epochs,
sketching the gravitational waveforms each produces and how they
encode source parameters.  It should be emphasized that this
characterization is rather crude.  The delineation between
``inspiral'' and ``merger'' in particular is not very clear cut,
especially when a binary's members are of comparable mass.  Despite
its crudeness, it is a very useful characterization for our purposes,
since parameterized waveforms exist for the inspiral and ringdown
epochs.

\subsection{Inspiral}

``Inspiral'' denotes the epoch in which the black holes are widely
separated from one another and slowly spiral together due to the
backreaction of gravitational waves upon the system.  These GWs are
well modeled (at least over most of this epoch) using the
post-Newtonian approximation to general relativity, roughly speaking
an expansion in inverse separation of the bodies; see Blanchet (2002)
and references therein for more detailed discussion.

The strongest harmonic of the inspiral waves has the form
\begin{equation}
h_{\rm insp}(t) = \frac{{\cal M}_z^{2/3}f(t)^{2/3}}{D_L}{\cal F}({\rm
angles}) \cos\left[\Phi(t)\right]\;,
\label{eq:insp_h}
\end{equation}
where ${\cal M}_z$ is the redshifted ``chirp mass'' (described in the
following paragraph), $D_L$ is the luminosity distance to the source,
$\Phi(t)$ is the accumulated GW phase, and $f(t) = (1/2\pi) d\Phi/dt$
is the GW frequency.  The function ${\cal F}({\rm angles})$ stands for
the rather complicated dependence of the signal on the position and
orientation angles of the source; see Cutler (1998) for a detailed
discussion of this dependence.

The phase function $\Phi$ depends very strongly upon the source's
chirp mass ${\cal M}$ (so called because it largely determines the
rate at which a binary's orbital frequency changes), and somewhat less
strongly upon the reduced mass $\mu$.  These mass parameters relate to
the masses which a binary's black holes would have in isolation, $m_1$
and $m_2$, by
\begin{equation}
{\cal M} = {(m_1 m_2)^{3/5}\over(m_1 + m_2)^{1/5}}\;,
\qquad
\mu = {m_1 m_2\over m_1 + m_2}\;.
\label{eq:insp_masses}
\end{equation}
A measurement of ${\cal M}$ and $\mu$ can thus determine the isolated
black hole masses, $m_1$ and $m_2$, fixing the initial mass $M_i = m_1
+ m_2$.  The phase $\Phi$ is the observable to which GW measurements
are most sensitive.  By measuring this phase and fitting to a model
(``template''), one can infer the source's physical parameters.  Since
${\cal M}$ most strongly impacts $\Phi$, it is the parameter which is
measured most accurately; $\mu$ does not impact $\Phi$ as strongly and
so is not measured as precisely.  The phase also depends on the spins
that the black holes would have in isolation, ${\bf S}_1$ and ${\bf
S}_2$.  We do not fully take this dependence into account; as we
discuss in {\S\ref{sec:conclude}} there is room for significant
improvement upon our analysis by doing so.

An interesting feature of GW measurements is that they do not actually
measure a binary's masses; rather, they measure {\it redshifted} mass
parameters.  This is because any quantity $m$ with the dimension of
mass enters the orbit evolution of the binary as a timescale $Gm/c^3$;
this timescale is then redshifted.  Hence, we measure ${\cal M}_z
\equiv (1 + z){\cal M}$ and $\mu_z \equiv (1 + z)\mu$.  As we discuss
in detail in {\S\ref{sec:measure}}, this subtlety has no impact on
this analysis, although it is crucial for many other studies (see,
e.g., Hughes 2002).

The ending of the inspiral is not perfectly well defined in all
circumstances, but corresponds roughly to when the members of the
binary are separated by a distance $r \sim 6GM/c^2$.  At this
separation, the system's evolution will certainly cease being slow and
adiabatic; the black holes will plunge towards one another and merge
into a single object.

\subsection{Ringdown}

The ringdown epoch consists of the last waves the system generates, as
the merged remnant of the coalescence relaxes to the quiescent Kerr
black hole state.  As it settles down, the distortions to this final
black hole can be decomposed into spheroidal modes, with
spherical-harmonic-like indices $l$ and $m$; the evolution of these
modes can then be treated using perturbation theory (Leaver 1985).
Each mode takes the form of a damped sinusoid.  Once the indices are
fixed, the frequency and damping time of these modes is determined by
the final mass and spin of the merged black hole:
\begin{equation}
h_{\rm ring}(t) \propto \exp(-\pi f_{\rm ring} t/Q) \cos(2\pi f_{\rm
ring} t + \varphi)\;.
\end{equation}
The indices most likely are fixed to $l = m = 2$.  This is a bar-like
mode oriented with the hole's spin.  Since a coalescing system has a
shape that nearly mimics this mode's shape, it should be
preferentially excited; also, it is more long-lived than other modes.
A good fit for this mode's frequency $f_{\rm ring}$ and quality factor
$Q$ is (Leaver 1985, Echeverria 1989)
\begin{eqnarray}
f_{\rm ring} &=& \frac{1}{2\pi(1 + z)M_f}\left[1 - 0.63(1 -
a)^{3/10}\right]\;,
\\
Q_{\rm ring} &=& 2(1 - a)^{-9/20}\;.
\end{eqnarray}
The parameter $(1+z)M_f$ is the redshifted mass of the final, merged
remnant of the binary; $a = |{\bf S}_f|/M_f^2$ is the dimensionless
Kerr spin parameter of the remnant, calculated from its spin vector
${\bf S}_f$.

Measurement of the ringdown waves thus makes it possible to determine
the final mass of the merged system.  We set the amplitude of this
mode by assuming that $1\%$ of the system's mass is converted into GWs
during the ringdown.  This is consistent with (or somewhat smaller
than) results seen in recent numerical simulations [see, e.g., Baker
et al.\ (2001) and references therein].  For the most part, our
results are only weakly dependent on this choice.  If the radiated
fraction is smaller than $1\%$ by more than a factor of a few, then in
some cases (particularly on the lower-mass end of the golden range we
discuss below), the final mass will not be determined well enough for
measurements to be golden.

The ringdown waves also allow us to determine the magnitude of the
merged system's spin angular momentum, $|{\bf S}_f|$.  We will not
take advantage of this feature in this analysis.  As we discuss in
{\S\ref{sec:conclude}}, measurement of $|{\bf S}_f|$ could play an
important role in an extension to our basic idea.

\subsection{Merger}

The ``merger'' epoch consists of all GWs generated between the
inspiral and ringdown.  Physically, this epoch describes the phase in
which the evolution ceases to be slow and adiabatic.  At least for
large mass ratio, the members of the binary encounter a dynamical
instability in their orbit and rapidly plunge together, merging into a
highly distorted object.  Because of the extreme strong field nature
of this epoch, neither a straightforward application of post-Newtonian
theory nor perturbation theory is very useful.

As discussed in the introduction, numerical and analytic work to
understand the merger epoch and the transition from inspiral to merger
is currently very active.  It is certainly to be hoped that, by the
time {\it LISA} flies, the GWs from this regime of coalescence will be
well understood for at least some important subsets of possible binary
black hole coalescences.  At present, the regime we call ``merger'' is
not modeled well enough to contribute to this analysis.  Indeed, there
may not even be a very useful delineation between ``inspiral'' and
``merger'' when $m_1 \simeq m_2$ --- the binary may smoothly evolve
from two widely separated bodies into a single object without
encountering any kind of instability or sharp transition (e.g.,
Pfeiffer, Cook, \& Teukolsky 2002).  However --- and this is one of
the major results of our analysis --- for the golden binaries, we will
be able to robustly constrain the GWs emitted from this regime solely
from the inspiral and ringdown waves, without needing detailed
modeling of the merger.

\section{Mass Measurement Accuracy}
\label{sec:measure}

Hughes (2002) has generated detailed Monte-Carlo simulations to assess
the precision with which {\it LISA} will measure BH masses in a
cosmological context.  Here we summarize the method and results, and
describe how the procedure was extended to estimate the precision on
mass deficit measurements for golden binary mergers.

\subsection{Mass deficit measurement accuracy: Formalism}

We estimate the accuracy with which {\it LISA} can measure binary
black hole parameters using a maximum likelihood parameter estimation
formalism originally developed in the context of GW measurements by
Finn (1992).  We begin with a parameterized model for the GW, written
schematically $h(\btheta)$, where $\btheta$ denotes a vector whose
components $\theta^a$ represent the masses of the black holes, their
spins, the position of the binary on the sky, etc.  From this model
and from a description of {\it LISA}'s sensitivity to GWs, we then
build the Fisher information matrix $\Gamma_{ab}$.  Schematically,
this matrix is given by
\begin{equation}
\Gamma_{ab} = 4{\rm Re}\int_0^\infty df\,
\frac{\partial_a{\tilde h}^*(f)\partial_b{\tilde h}(f)}{S_h(f)}\;.
\end{equation}
The function $S_h(f)$ is the detector's noise spectral density, which
we discuss in more detail below; $\tilde h(f)$ is the Fourier
transform of $h(t)$, and $\partial_a \equiv
\partial/\partial\theta^a$.  The $*$ superscript denotes complex
conjugate.  For detailed discussion of how we compute $\Gamma_{ab}$,
see Hughes (2002); that paper in turn relies heavily upon the
discussion in Cutler (1998), Poisson \& Will (1995), Cutler \&
Flanagan (1994), and Finn \& Chernoff (1993).  Note in particular that
we describe the {\it LISA} response using the formalism developed by
Cutler, which synthesizes a pair of ``effective detectors'' from the
data on the three {\it LISA} arms; see Cutler (1998) for details.

The inverse Fisher matrix gives us the accuracy with which we expect
to measure the parameters $\theta^a$:
\begin{equation}
\Sigma^{ab} = \langle\delta\theta^a \delta\theta^b\rangle
= (\bGamma^{-1})^{ab}\;.
\end{equation}
In this equation, $\delta\theta^a$ is the measurement error in
$\theta^a$; the angle brackets denote an ensemble average over all
possible realizations of noise produced by the detector.  Thus, the
diagonal components, $\Sigma^{aa} = \langle(\delta\theta^a)^2\rangle$,
represent the expected squared errors in a measurement of $\theta^a$;
off-diagonal components describe correlations between parameters.

For our purposes, the most important result is that for golden
binaries, we measure the chirp mass $\cal M$, the reduced mass $\mu$,
and the final mass of the merged system $M_f$.  From these, we
construct the GW mass deficit $M_{\rm GW}$.  As discussed in the
previous section, we would actually measure {\it redshifted} mass
parameters, $(1 + z){\cal M}$, $(1 + z)\mu$, $(1 + z)M_f$, and $(1 +
z)M_{\rm GW}$.  The redshift is irrelevant to us here: since all of
our masses are measured at the same $z$, the factor $1 + z$ is an
uninteresting overall rescaling.  We will present all of our error
estimates in dimensionless form ($\delta{\cal M}/{\cal M}$, $\delta
M_{\rm GW}/M$, etc.), so that this dependence scales out.  This has
the important effect that uncertainties in cosmological parameters do
not effect our analysis.

Since $\cal M$ and $\mu$ depend upon the masses of the two black holes
which constitute the binary, we infer the initial total mass of the
system $M_i$ from $\cal M$ and $\mu$: using the definition ${\cal M} =
\mu^{3/5}M_i^{2/5}$, we have
\begin{equation}
M_i = \mu^{-3/2}{\cal M}^{5/2}\;.
\label{eq:M_i}
\end{equation}
From the difference between $M_i$ and $M_f$, we then infer how much of
the system's mass was radiated in GWs:
\begin{equation}
M_{\rm GW} \equiv M_i - M_f\;.
\end{equation}
Our goal is to assess what errors $\delta M_{\rm GW}$ are likely.  The
mass loss is itself a fairly small quantity --- we expect $M_{\rm
GW}/M \lesssim 10 - 20\%$ at best.  Clearly, the ratio $\delta M_{\rm
GW}/M$ must be smaller than this if the measurement of $M_{\rm GW}$ is
to have any meaning.  We now relate $\delta M_{\rm GW}$ to the errors
$\delta\cal M$, $\delta\mu$, and $\delta M_f$ estimated by our
Monte-Carlo calculations.

Because the inspiral and ringdown epochs are likely to be analyzed
separately, our estimates of $M_i$ and $M_f$ should be statistically
independent of one another.  The errors in these quantities thus
combine in quadrature for the error $\delta M_{\rm GW}$:
\begin{equation}
\delta M_{\rm GW}^2 = \delta M_i^2 + \delta M_f^2\;.
\end{equation}
Since $M_i \simeq M_f \simeq M$, it is convenient to divide by the
mass of the system to write this as
\begin{equation}
\left(\frac{\delta M_{\rm GW}}{M}\right)^2 \simeq \left(\frac{\delta
M_i}{M_i}\right)^2 + \left(\frac{\delta M_f}{M_f}\right)^2\;.
\end{equation}
This allows us to scale out the uninteresting $(1 + z)$ factor
attached to all masses.  Also, our Monte-Carlo code's estimates of
error in any mass parameter $m$ are given in the form $\delta\ln(m) =
\delta m/m$; see Hughes (2002) for details.

Using Eq.\ (\ref{eq:M_i}), we find the following relation for the
error $\delta M_i$ in terms of errors in the chirp and reduced masses:
\begin{eqnarray}
\left(\frac{\delta M_i}{M_i}\right)^2 &\simeq& \left(\frac{\partial
M_i}{\partial\mu}\right)^2 \left(\frac{\delta\mu}{M_i}\right)^2 +
\left(\frac{\partial M_i}{\partial\cal M}\right)^2
\left(\frac{\delta{\cal M}}{M_i}\right)^2 \nonumber\\
&\simeq& \frac{9}{4}\left(\frac{\delta\mu}{\mu}\right)^2 +
\frac{25}{4}\left(\frac{\delta\cal M}{\cal M}\right)^2\;.
\end{eqnarray}

Our final expression for the error in $M_{\rm GW}$, normalized to the
total mass of the binary, is thus
\begin{equation}
\frac{\delta M_{\rm GW}}{M} \simeq \left[
\frac{9}{4}\left(\frac{\delta\mu}{\mu}\right)^2 +
\frac{25}{4}\left(\frac{\delta\cal M}{\cal M}\right)^2 +
\left(\frac{\delta M_f}{M_f}\right)^2\right]^{1/2}\;.
\end{equation}
We build the likely distributions of this error parameter through
Monte-Carlo simulation.  We choose masses for the binaries' members,
we choose the binaries' redshift, and we then randomly distribute such
binaries in sky position, source orientation, and final merger time.

Our procedure for doing this analysis is nearly identical to that
described in \S4 of Hughes (2002), so we will not discuss it in detail
here.  We do note two important differences between Hughes (2002) and
this analysis.  First, we include here the effect of mass ratio.
Hughes (2002) focused, for simplicity, on equal mass binaries.  For a
fair assessment of the likely importance of golden binaries, inclusion
of mass ratio is necessary.  Second, we have updated the description
of the {\it LISA} noise $S_h(f)$; we now use the sensitivity described
by Barack and Cutler (2004) [their Eqs.\ (48)--(54)].  This corrects
some errors in Hughes (2002) and is in accord with current plans for
the {\it LISA} mission.

We make one modification to Barack and Cutler's noise curve: we assume
that the detector's response is only well-understood above some
fiducial frequency $f_{\rm low}$, so that we ignore all GWs that
radiate at $f < f_{\rm low}$.  During the inspiral, a binary's orbital
frequency monotonically increases as its black holes gradually fall
towards one another.  By ignoring waves with $f < f_{\rm low}$, we are
effectively saying that a signal becomes visible to {\it LISA} when
its GW frequency becomes greater than $f_{\rm low}$.  An appropriate
value for $f_{\rm low}$ is a matter of some debate; physically, we
expect that $f_{\rm low} \simeq 2\pi/T_{\rm DF}$, where $T_{\rm DF}$
is the maximum time over which the {\it LISA} spacecraft can maintain
``drag-free'' orbital motion.  Much work on {\it LISA} science assumes
that $f_{\rm low} \simeq 10^{-4}\, {\rm Hz}$; the {\it LISA} science
team is currently investigating whether the noise can be characterized
down to $f_{\rm low} \simeq 3\times10^{-5}\,{\rm Hz}$.  Lower values
of $f_{\rm low}$ typically lead to improved accuracy for parameters of
the inspiral waveform, since more phase is measured (see Hughes \&
Holz 2003, Figure 2); this can be particularly important for higher
black hole masses, since GW frequencies are inversely proportional to
mass.

In the results that we present here, we have set $f_{\rm low} =
3\times10^{-5}\,{\rm Hz}$, assuming (perhaps optimistically) that the
instrument's noise will be well-characterized to relatively low
frequencies.  We have also examined golden binaries for the choice
$f_{\rm low} = 10^{-4}\,{\rm Hz}$.  Though we do not discuss these
results in detail, the impact of this choice can be simply summarized:
the accuracy with which both the chirp mass $\cal M$ and the reduced
mass $\mu$ are measured is degraded.  The degradation in $\cal M$ is
not important --- $\cal M$ is determined so precisely that it is
largely irrelevant in setting the accuracy with which $M_{\rm GW}$ is
determined.  We typically find $\delta{\cal M}/{\cal M} \sim 10^{-3} -
10^{-4}$; this phenomenal accuracy is because the chirp mass largely
determines the number of orbits which are radiated in the detector's
band.  A phase coherent measurement is very sensitive to this number,
and thus determines ${\cal M}$ very precisely.

By contrast, the degradation in $\mu$ has a very large effect.  In the
best cases, we find that $\mu$ is determined with an accuracy of a
fraction of a percent to a few percent; in most cases, it dominates
the error budget for $M_{\rm GW}$.  When $f_{\rm low} = 10^{-4}\,{\rm
Hz}$ is used instead of $3\times10^{-5}\,{\rm Hz}$, the error in $\mu$
typically increases by a factor $\sim 3 - 10$.  This sharply restricts
the mass range that can be considered golden.  It is likely that this
tendency can be offset by taking into account, as in Vecchio (2004),
modulations due to spin precessions which break certain degeneracies
and allow $\mu$ to be determined much more accurately.

\subsection{Mass deficit measurement accuracy: Results}

Our Monte-Carlo distributions for $\delta M_{\rm GW}$ show that golden
binaries exist in a somewhat narrow band of mass, typically from $M_z
\sim {\rm several}\times 10^{5}\, M_\odot$ to $M_z \sim {\rm
several}\times 10^6\,M_\odot$, where $M_z = (1 + z)M$, and $M$ is the
binary's total mass.  Binaries more massive than this do not radiate
in band long enough to determine the system's reduced mass with
sufficient accuracy to be golden; less massive binaries do not have a
ringdown signal that is loud enough to determine the system's final
mass with sufficient accuracy.  The upper and lower bounds on this
mass window evolve with redshift, so that the mass window shrinks as
we move to higher $z$.

Figures {\ref{fig:one}--{\ref{fig:five}} show representative
probability distributions for the measurement error $\delta M_{\rm
GW}$ as a function of binary masses and redshift.  Figures
{\ref{fig:one}} and {\ref{fig:two}} show results for small redshift,
$z = 0.5$.  Although we do not expect a very interesting event rate at
this redshift, this result gives a ``best case'' illustration of how
well these measurements can be performed in principle --- events at $z
= 0.5$ are so strong that mass parameters are determined with very
high precision.  Figure {\ref{fig:one}} illustrates the ``goldenness''
of the range $2\times 10^5\,M_\odot \lesssim M \lesssim 2\times
10^6\,M_\odot$ at this redshift.  The most likely error (peak of the
distribution) $\delta M_{\rm GW}/M$ is about $2-3\%$ at the
extrema of the range ($m_1 = m_2 = 10^5\,M_\odot$ and $m_1 = m_2 =
10^6\,M_\odot$), and goes down to $\delta M_{\rm GW}/M \simeq 1\%$ in
the best cases.

Although these distributions show a peak representing excellent
measurement precision, it must be emphasized that each distribution
also has a rather extended, large error tail.  These tails are mostly
due to a subset of MBH binaries in the Monte Carlo realizations that
have a short inspiral time before coalescence at the time of
detection.

As we move to higher and lower masses, the peak of the error
distribution shifts outside of the golden range.  Figure
{\ref{fig:two}} illustrates the error distributions at $z = 0.5$ just
above and just below the masses shown in Fig.\ {\ref{fig:one}}.
Although there is still a good probability of measuring a golden event
at these masses, they are no longer the most likely events.  As we move
to even higher and lower masses, the probability of a golden event
decreases more rapidly.

The rule of thumb we learn from Figs.\ {\ref{fig:one}} and
{\ref{fig:two}} is that the most probable mass range for golden
binaries are those with a redshifted mass ${\rm several}\times10^5\,
M_\odot \lesssim M_z \lesssim {\rm a\ few}\times10^6\,M_\odot$.  As
Figs.\ {\ref{fig:three}} -- {\ref{fig:five}} illustrate, this pattern
continues out to $z \sim 2 - 4$.  At $z = 1$, most of the merger
events in this mass range will be golden; see Fig.\ {\ref{fig:three}}.
Beyond this range, the probability that an event is golden begins to
decrease, largely due to the weakness of signals at this distance,
which makes determining the final mass difficult.  Fig.\
{\ref{fig:four}} shows the distribution of errors at $z = 2$.  Many
events are still golden at this redshift, though the proportion is
decreasing.  By the time we reach $z = 4$ (Fig.\ {\ref{fig:five}}),
the proportion of golden events has fallen quite a bit.

By considering the peaks of the many error distributions that we have
computed, we determine a typical range of masses, as a function of
redshift, for which binaries are golden (i.e. with a distribution of
$\delta M_{\rm GW}/M$ peaking at $5\%$ or less).  Beyond $z \sim 4$,
{\it LISA} cannot determine $M_{\rm GW}$ well enough for there to be a
reasonable probability of measuring a golden merger; but, a fairly
large range of masses are golden at lower redshifts. We find that a
simple mass-redshift prescription reproduces our results reasonably
well, even though it does not capture all of the information contained
in error distributions such as those shown in
Figs.~\ref{fig:one}--\ref{fig:five}. For a black hole binary to be
golden, the largest mass black hole, $m_1$, should satisfy the
constraint
\begin{equation} \label{eq:cons1}
5 < \log \left( \frac{m_1}{M_\odot} \right) < 5 +1.3 \times 
\left( \frac{4-z}{3.5} \right),
\end{equation}
which results in no binary being classified as golden beyond $z=4$. In
addition, the lowest mass black hole, $m_2$, should satisfy the constraint
\begin{equation} \label{eq:cons2}
0.3 \leq q \equiv m_2/m_1 \leq 1.
\end{equation}
This constraint on the binary mass ratio is partly the result of our
error analysis (binaries with small mass ratios are often not golden).
It is also motivated by the expectation that radiative mass losses,
$M_{\rm GW}/M$, are small in mergers with fairly unequal masses. While
losses can perhaps reach $\sim 10$-$20 \%$ in a binary with $q \sim 1$
(depending on the spins and relative orientations of the merging BHs),
an expected $q^2$ scaling for the losses makes it unlikely that
$M_{\rm GW}$ can be measured in binaries with small mass ratios (see,
e.g., Menou \& Haiman 2004 and references therein for a discussion of
these scalings). We have chosen $q=0.3$ as a limit below which losses
(likely $< 1$-$2 \%$) become very difficult to measure with {\it
LISA}.

Having demonstrated the possibility of radiative mass loss
measurements with {\it LISA}, it remains to be seen whether enough
golden binary mergers satisfying the above constraints will occur for
{\it LISA} to actually carry out any such measurement.  Using specific
population models, we argue in the next section that the event rate of
golden binary mergers is rather low, but nonetheless high enough that
$M_{\rm GW}$ measurements are plausible for a nominal 3--year {\it
LISA} mission lifetime.

\section{Population Models and Event Rates}
\label{sec:rates}

At present, too little is known about the characteristics of the
population of MBHs at cosmological distances to make robust merger
rate predictions for {\it LISA}. Here, our goal is to show that one
may reasonably expect {\it LISA} to make some radiative mass loss
measurements from golden binaries. To do so, we rely on pre-existing
population models, which were described at length in Menou et al.\
(2001) and Menou (2003).  Though simple, these models capture the
essential galactic merger process which is at the origin of MBH
coalescences. Let us summarize here their main characteristics.

The basic ingredient of these models is a ``merger tree'' describing
the cosmic evolution of dark matter halos in a concordance
$\Lambda$--CDM cosmology (Press \& Schechter 1974; Lacey \& Cole 1993,
1994). Unless otherwise specified, all model parameters adopted here
are identical to those in Menou et al. (2001). By identifying dark
matter halos with individual galaxies and by populating these galaxies
with central MBHs, it is possible to follow the cosmic evolution of
galactic mergers and trace the occurrence of MBH coalescences through
cosmic times.

While this simple description of galactic and MBH mergers has received
some basic observational support (see, e.g., Komossa et al.\ 2003), it
also hides a number of complications and related uncertainties.  Even
though dynamical studies indicate that nearly all nearby massive
galaxies harbor central MBHs (Magorrian et al.\ 1998), it is unclear
how frequently these MBHs are present in galaxies at higher redshifts
(Menou et al.\ 2001; Volonteri, Haardt \& Madau 2003). In addition, the
timescale on which two MBHs coalesce, within a merged galactic
remnant, is not well known. Dynamical friction is the initial
mechanism bringing the two MBHs together, but the timescale on which
it operates becomes increasingly long for small mass ratios (Yu
2002). The second mechanism bringing the two MBHs together, which
involves three-body encounters with stars on very low angular momentum
orbits (in the ``loss cone''), may be rather inefficient (see, e.g.,
Begelman, Blandford \& Rees 1980; Milosavljevic \& Merritt 2003). On
the other hand, interaction between the two MBHs and a gaseous
component may help accelerate their coalescence substantially
(e.g. Gould \& Rix 2000; Escala et al.\ 2004).  To assess the
likelihood of {\it LISA} performing accurate mass loss measurements
over its mission lifetime, we will focus on an optimistic scenario in
which it is assumed that MBHs coalesce efficiently following the
merger of their host galaxies. We will then comment on the
consequences of relaxing this assumption for our results.

We must specify the mass properties of the cosmological population of
MBHs in our models. In recent years, spectacular progress has been
made in characterizing the masses of MBHs at the centers of nearby
galaxies. A tight correlation between BH mass and velocity dispersion
of the galactic spheroidal component has been established via detailed
dynamical studies (Gebhardt et al.\ 2000; Ferrarese \& Merritt 2000;
Tremaine et al.\ 2002). A relation between BH mass and galactic mass
has subsequently been proposed (Ferrarese 2002). Evidence that this
link between central MBHs and their host galaxies is already in place
at redshifts $z \sim 3$ has also been presented (Shields et al.\
2003), although more recent work suggests that this link may evolve
rather significantly (Treu, Malkan, \& Blandford 2004).

In view of these developments, we assume that the following relation
between the galactic mass and the BH mass, $M_{\rm bh}$, is satisfied
at all times in our population models (Ferrarese 2002; Wyithe \& Loeb
2004):
\begin{equation} \label{eq:mass}
M_{\rm bh}= 10^9 M_\odot \left( \frac{M_{\rm halo}}{1.5 \times 10^{12}
M_\odot}\right)^{5/3} \left( \frac{1+z}{7} \right)^{5/2},
\end{equation}
where $M_{\rm halo}$ is the mass of the dark matter halo associated
with each galaxy.  This relation may result from radiative or
mechanical feedback by the quasar during active periods of BH mass
accretion (e.g. Silk \& Rees 1998; Murray, Quataert \& Thompson
2004). We do not explicitly model BH accretion in our models, but it
is done implicitly by enforcing the above relation for all MBHs, at
all epochs.

Equipped with this population model, it is now possible to calculate
the merger rate of MBH binaries, including the subset of binaries
satisfying the ``golden'' conditions given in Eqs.~(\ref{eq:cons1})
and~(\ref{eq:cons2}). We consider two separate models, starting at
$z=5$. In one model, we assume that MBHs are abundant in the sense
that all potential host galaxies at $z=5$ do harbor a MBH. In a second
model, we assume that MBHs are $\sim 30$ times rarer, by confining
them to the $3 \%$ most massive galaxies described by the merger tree
at $z=5$. This was found to be about the lowest occupation fraction at
that redshift which remains consistent with the observational
requirement of ubiquitous MBHs in massive galaxies at $z=0$ by Menou
et al.\ (2001). Apart from assigning a mass to each MBH
(Eq.~[\ref{eq:mass}]), our models are thus identical to two of the
three models discussed in detail in Menou et al.\ (2001).

Figure~\ref{fig:six} shows the event rates for golden binary mergers
predicted by these two models (after averaging over ten independent
realizations of the merger tree). The dashed line corresponds to the
model with abundant MBHs and the dotted line to the model with rare
MBHs. For comparison, we also show the total event rate in the model
with rare MBHs as a solid line (same as in Fig.~2b of Menou et al.\
2001). As expected from our ``goldenness'' definition, the rate of
golden binary mergers falls precipitously at $z \sim 3.5$-$4$. In both
models, it peaks around $z \sim 2$-$3$, where it reaches $\sim 1\%$ of
the total merger rate in the model with abundant MBHs (not shown) and
$\sim 10\%$ of the total merger rate in the model with rare MBHs
(compare solid and dotted lines).

In the model with abundant MBHs, the total number of golden events,
integrated over all redshifts, adds up to $N_{\rm gold} \sim 5$ for a
3-year {\it LISA} mission lifetime. It reduces to $N_{\rm gold} \sim
1$ in the model with rare MBHs. These numbers are encouraging because
they suggest that {\it LISA} can reasonably be expected to perform one
or several accurate radiative mass loss measurements. Of course, one
should also keep in mind that our models are partly optimistic in
assuming efficient MBH coalescences after galactic mergers. The lower
efficiency of dynamical friction at low mass ratios should not affect
much our predictions for golden binary mergers because we have
restricted golden binaries to have mass ratios $q \ge 0.3$
(Eq.~[\ref{eq:cons2}]). But, if MBH coalescences are much less
efficient than assumed here, because of the difficulty in replenishing
``loss cones'' and of the absence of an accelerating gaseous
component, merger rates for binaries of all types could be well below
our predictions. This could make the occurrence of golden events
rather unlikely. In that case, it may also be the case that the total
event rate for {\it LISA} would turn out to be a disappointingly low
number.  On the other hand, it should be borne in mind that we have
been somewhat conservative in our estimates of how precisely masses
can be measured.  The range of binary mass that can be considered
golden is probably somewhat broader than we have assumed, which may
offset somewhat any loss in the rate of golden events.

\section{Discussion and Conclusion}
\label{sec:conclude}

Within a certain mass range, {\it LISA} will be able to measure the
GWs that come from all epochs of the binary black hole coalescence
process.  By precisely measuring the initial and final masses of the
system, {\it LISA} will determine how much of the system's mass is
radiated away.  Such a measurement will constitute an extremely
interesting and robust probe of strong-field general relativity.

As we have emphasized, a particularly interesting feature of these
``golden'' binaries is that the mass deficit can be determined without
a detailed model of the highly dynamical, strong field plunge and
merger epoch.  Although it is almost certain that our theoretical
understanding of this epoch will be much improved by the time {\it
LISA} flies, we should not be surprised if substantial uncertainties
remain in the {\it LISA} era.  A simple and robust measure of
strong-field physics such as $M_{\rm GW}$ can play a particularly
important role in developing a phenomenology of binary black hole
merger physics.  Such an observable may be useful as a kind of global
calibration on the process, even if the waveform models for the
plunge/merger regime can not be computed with good accuracy.

We also emphasize again that our estimates for the accuracy with which
$M_{\rm GW}$ will be measured are, in all likelihood, somewhat
pessimistic.  Especially for the larger masses, much of our error
$\delta M_{\rm GW}$ is due to a (relatively) poorly determined value
of the binary's reduced mass $\mu$.  Parameter measurement analyses of
post-Newtonian waveforms have shown such large uncertainties are
typically due to correlations between $\mu$ and so-called
``spin-orbit'' and ``spin-spin'' parameters --- parameters which
depend upon couplings between the binary's orbital angular momentum
and black hole spin vectors (Cutler \& Flanagan 1993, Poisson \& Will
1995).  Vecchio (2004) has shown that these correlations can be broken
by carefully taking into account spin-induced modulations in the
orbital motion of the binary and their influence upon the
gravitational waveform.  In some cases, the error $\delta\mu$ can be
reduced by an order of magnitude or better.  The range of golden
binaries thus likely extends to somewhat larger masses than we have
estimated, improving the likelihood that {\it LISA} will measure these
events.  This effect may also allow binaries to be measured with
``golden'' precision even if our description of {\it LISA}'s
low-frequency noise turns out to be too optimistic.

Taking into account spin-induced precessions not only reduces the
error with which $\mu$ is measured; it also makes possible a
determination of the binary's individual black hole spins, ${\bf S}_1$
and ${\bf S}_2$.  As we mentioned in {\S\ref{sec:model}}, ringdown
waves allow us to determine the merged remnant's spin magnitude $|{\bf
S}_f|$ as well as its mass; in fact, spin and mass are measured with
essentially the same relative precision (Finn 1992).  This suggests
another extremely interesting measurement that might be possible with
these binaries: demonstrating that the initial and final
configurations of the system satisfy the {\it area theorem} (Hawking
1971).  Although the mass of the system must decrease because of
radiative losses,
\begin{equation}
M_i = M_1 + M_2 > M_f\;,
\end{equation}
the area of all event horizons in the system must {\it increase}
during the coalescence:
\begin{equation}
A_i = A_1 + A_2 < A_f\;.
\end{equation}
The area of a black hole's event horizon is determined totally by its
mass and angular momentum:
\begin{equation}
A = 8\pi M^2\left(1 + \sqrt{1 - a^2}\right)\;.
\end{equation}
(Here, $a = |{\bf S}|/M^2$ is the dimensionless Kerr parameter.)  This
measurement would demonstrate observationally a fundamental law of
black hole physics.

\acknowledgments

SAH thanks Sterl Phinney, Curt Cutler, and Alberto Vecchio for helpful
discussions.  He also thanks Kip Thorne for a helpful conversation
which led us to consider the possibility of golden binaries testing
the area theorem.  KM thanks Zoltan Haiman for the use of output data
from his merger tree code. This work was supported at MIT by NASA
Grant NAG5-12906 and by NSF Grant PHY-0244424.

\clearpage

\begin{figure}
\plotone{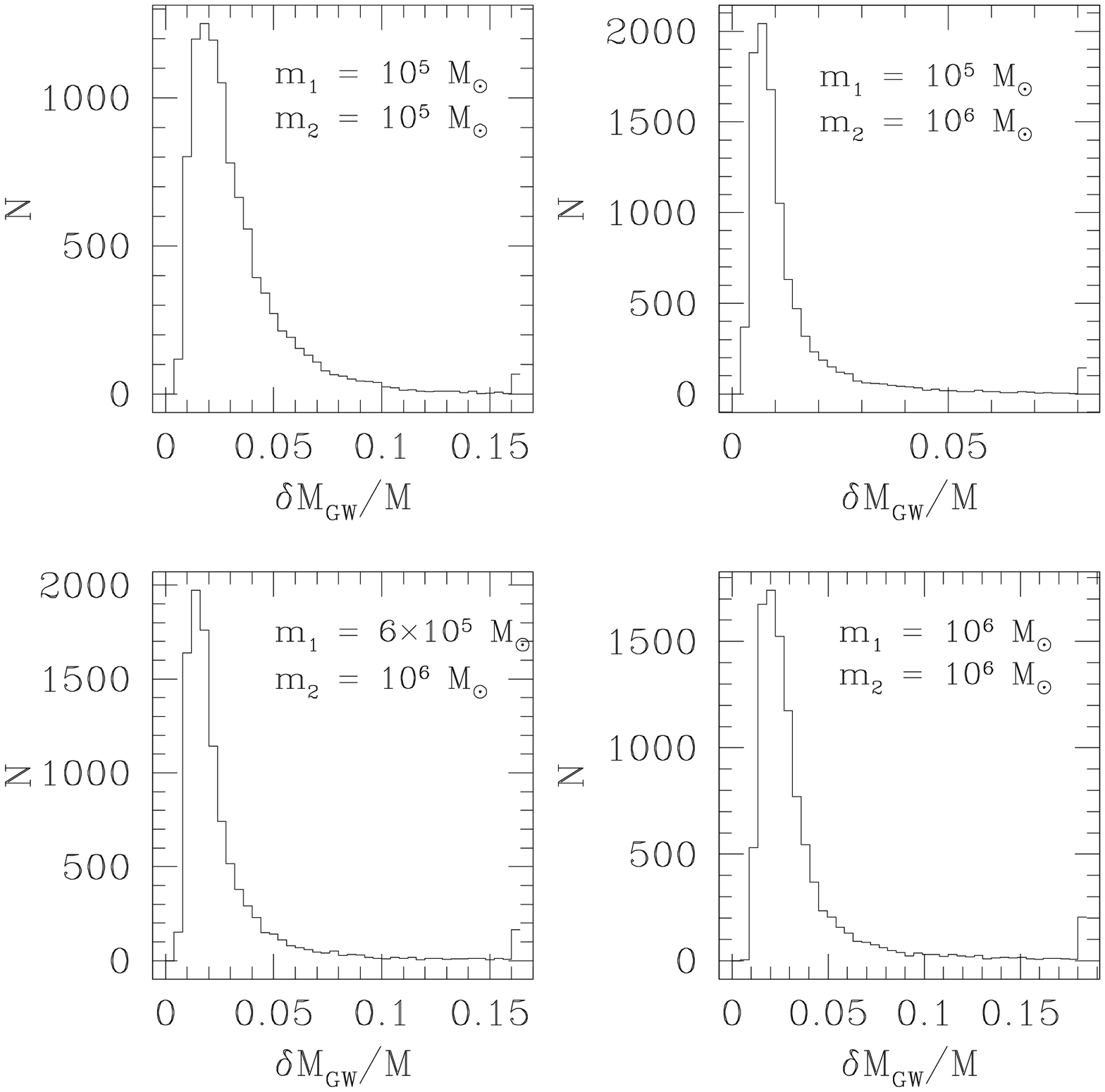}
\caption{Distributions of measurement uncertainty, $\delta M_{\rm
GW}/M$ (where $M$ is total binary mass), at $z = 0.5$.  Each panel
represents different choices for the binary masses.  As discussed in
the text, we randomly populate the sky with 10,000 of these binaries,
randomly distributing their orientation vectors and the time at which
they coalesce.  At this redshift, binaries with these masses are
solidly golden: the measurement error distributions peak at relative
error of $1\% - 3\%$.  We should have little difficulty measuring the
mass change due to radiative losses if nature provides us with
binaries at these masses and this redshift.
\label{fig:one}}
\end{figure}

\begin{figure}
\plotone{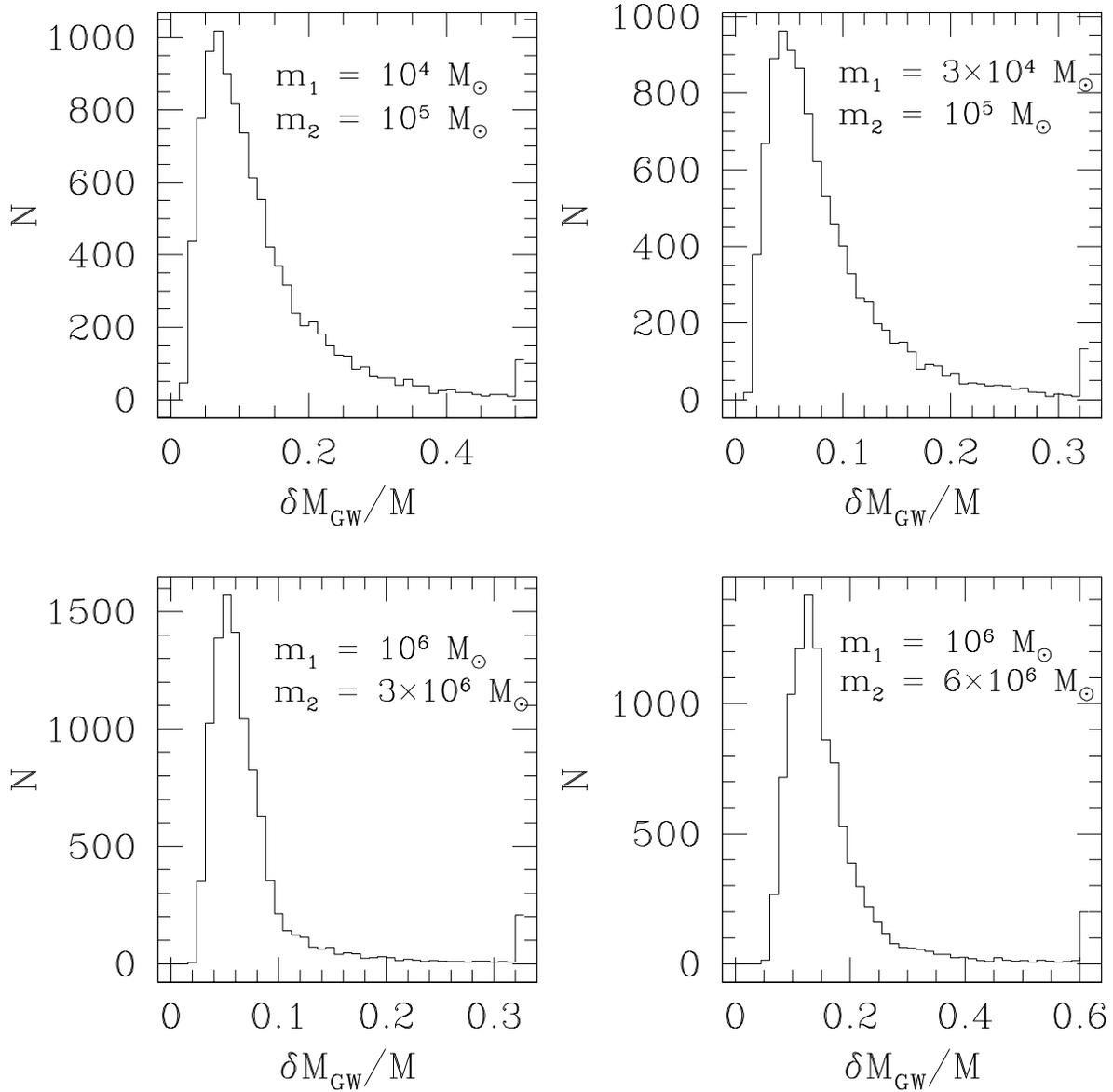}
\caption{Distributions of measurement uncertainty at $z = 0.5$.  We
show here the distribution of measurement uncertainty for binaries
just outside the range shown in Fig.\ {\ref{fig:one}}.  As we make the
total mass smaller, the final mass becomes more poorly determined due
to the weakness of the ringdown signal.  As we make the total mass
larger, the reduced mass becomes more poorly determined as less time
is spent in {\it LISA}'s band.  These distributions peak at a relative
error of $5\% - 8\%$; in each case, there is substantial probability
for $\delta M_{\rm GW}/M \lesssim 3\%$, though it is not the most
likely outcome.  There is thus a reasonable likelihood of golden
events in these cases, though with much less certainty than in the
mass range shown in Fig.\ {\ref{fig:one}}.  As we decrease and
increase the masses further, we find that the likelihood of golden
events dwindles very rapidly.
\label{fig:two}}
\end{figure}

\begin{figure}
\plotone{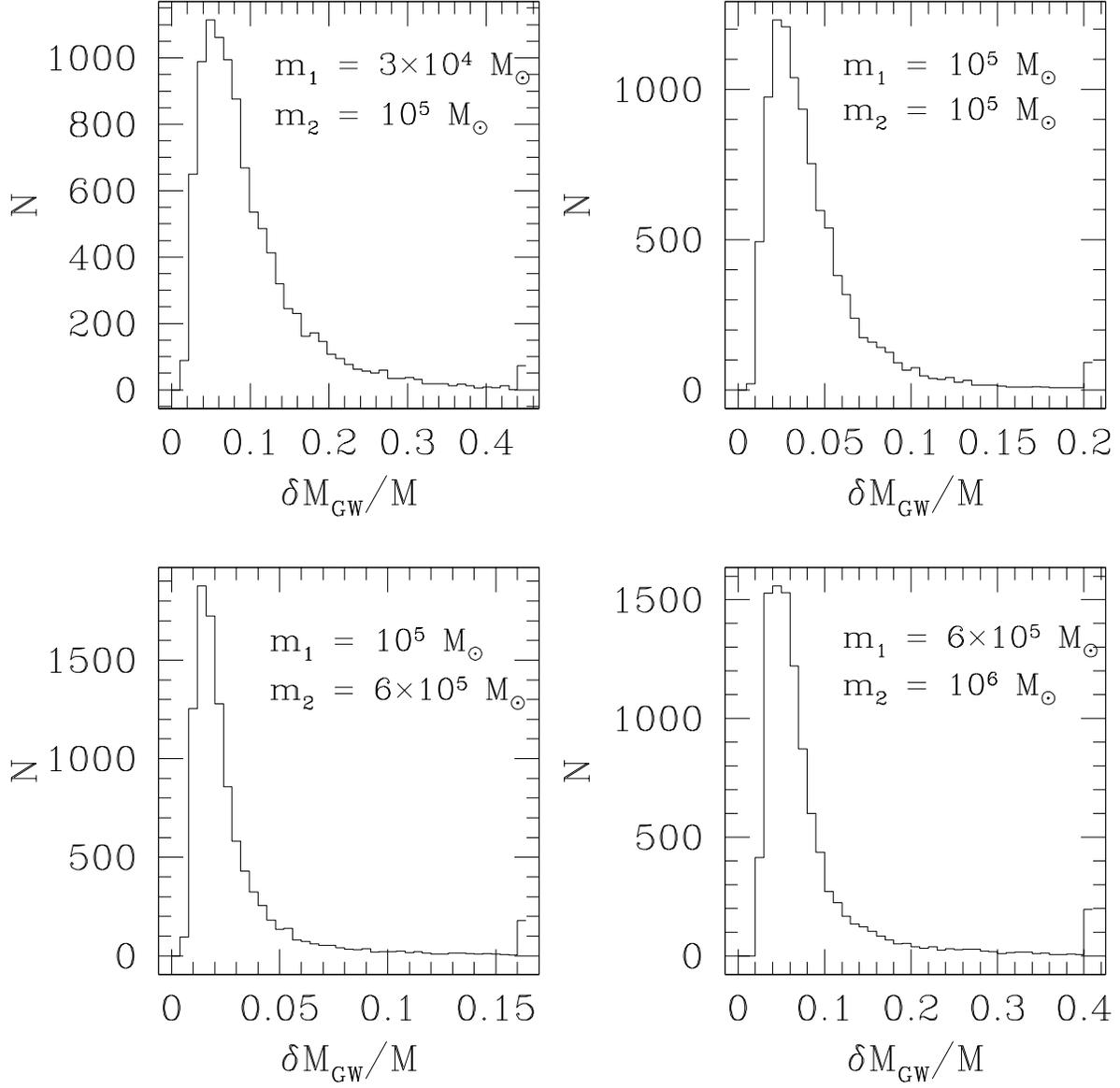}
\caption{Distributions of measurement uncertainty at $z = 1$.  The
mass range containing golden binaries gets smaller, as all signals
become weaker at this larger distance, and systematically shifts to
somewhat smaller total mass.  This shift is in accord with the fact
that GW measurements are sensitive to redshifted mass, $m_z = (1 +
z)m$, for any mass parameter $m$.  Many events in the mass range
shown here are likely to be golden.  The likelihood of golden events
rapidly falls away as we move outside of this mass range.
\label{fig:three}}
\end{figure}

\begin{figure}
\plotone{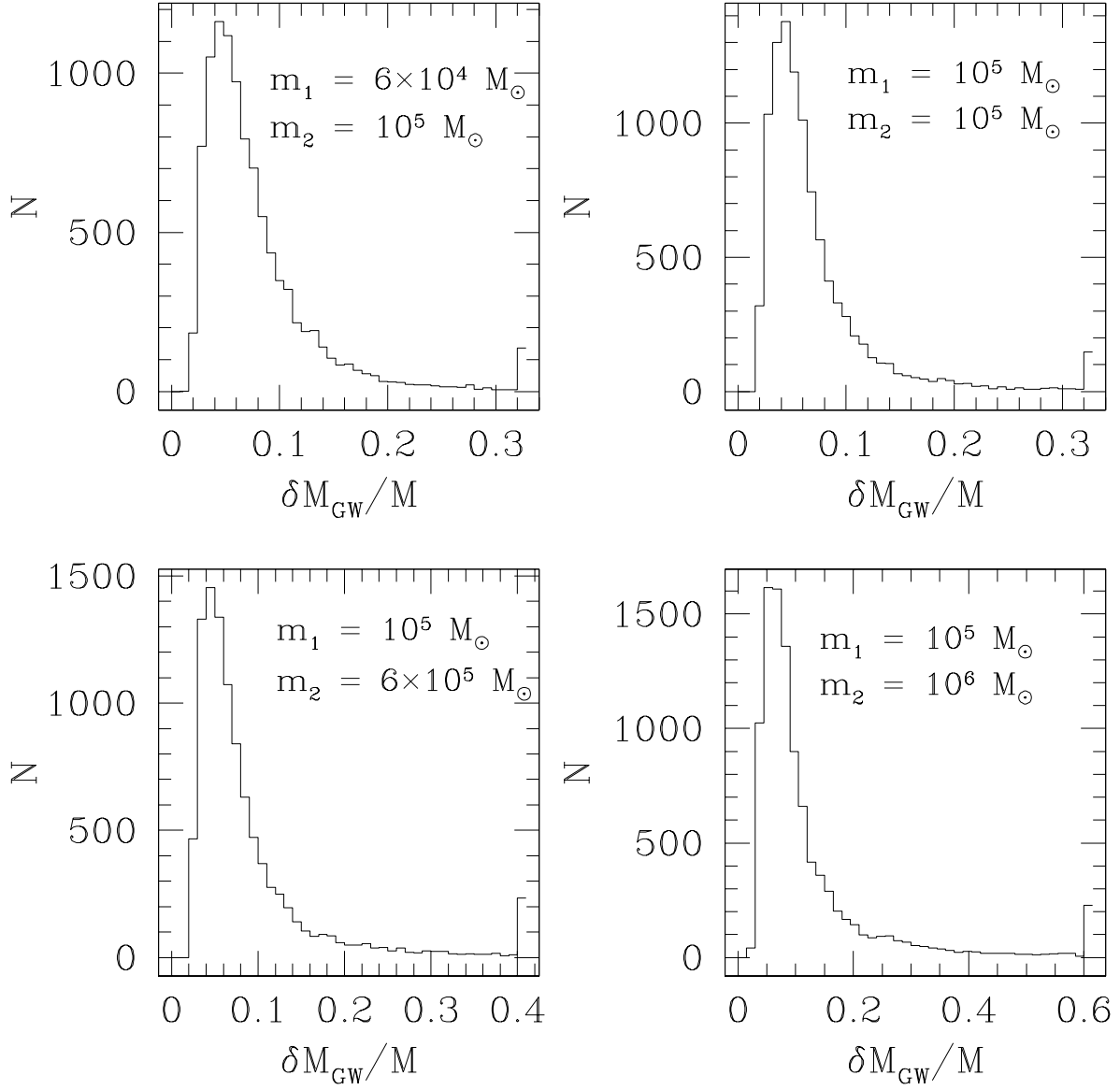}
\caption{Distributions of measurement uncertainty at $z = 2$.  We
continue to see the mass range containing golden binaries getting
smaller.  Events with total mass $M \sim {\rm several} \times
10^5\,M_\odot$ are most likely to be golden at this redshift.
\label{fig:four}}
\end{figure}

\begin{figure}
\plotone{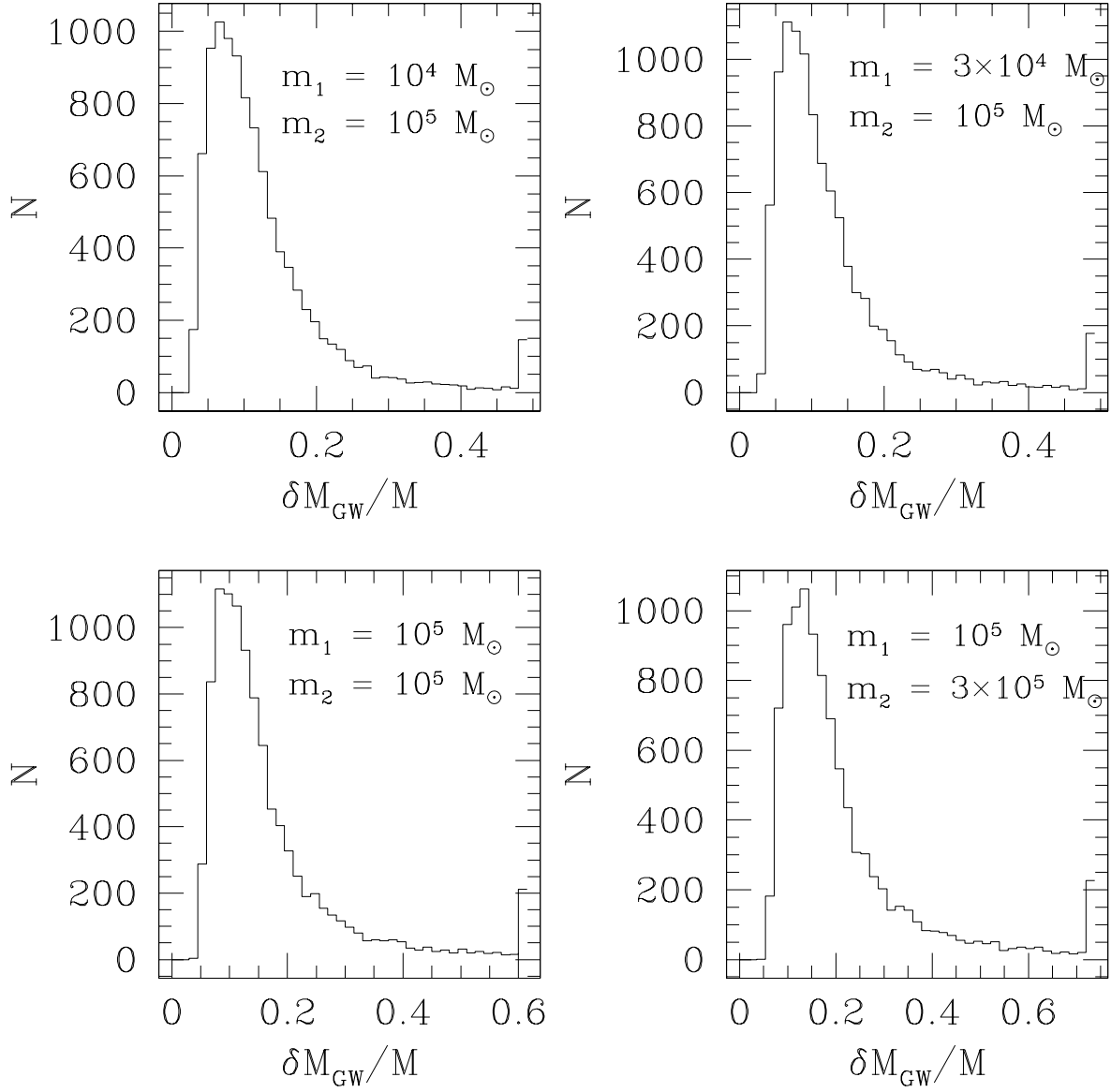}
\caption{Distributions of measurement uncertainty at $z = 4$.  As we
go to these relatively large redshifts, the precision of parameter
determination degrades enough that golden events become far less
likely --- we typically have $\delta M_{\rm GW}/M \lesssim 10\%$, not
quite good enough to measure with confidence the radiative mass
deficit.  Only relatively rare events at this redshift and in this
mass regime are expected to be golden.
\label{fig:five}}
\end{figure}

\clearpage

\begin{figure}
\plotone{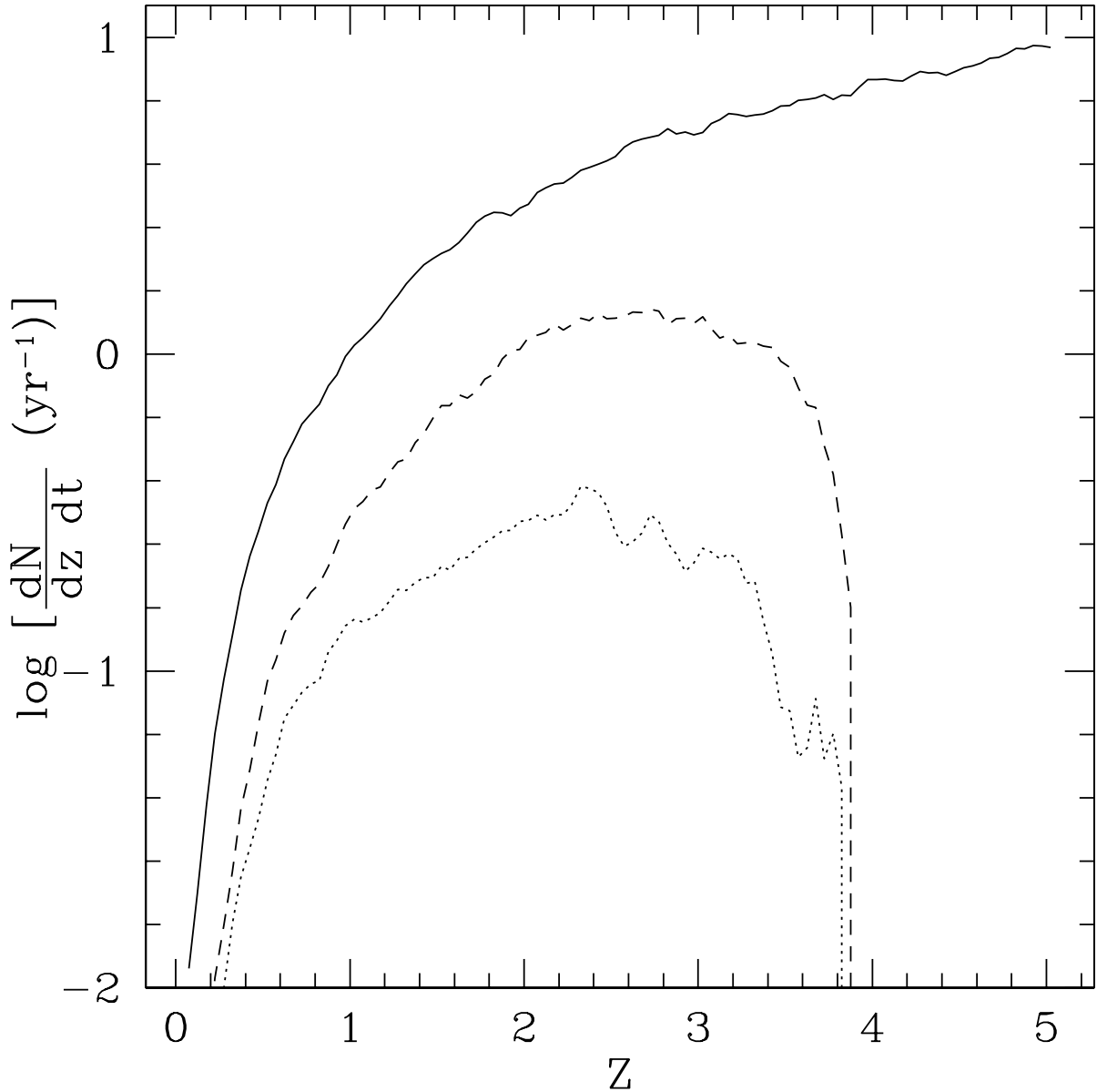}
\caption{Golden binary merger rates for {\it LISA}, per year and per
unit redshift, as a function of redshift, $z$. The dashed and dotted
lines correspond to models with abundant and rare populations of MBHs,
respectively (see text for details). For comparison, the total rate of
MBH binary mergers in the model with rare MBHs is also shown as a
solid line.  All the rates were smoothed over $\delta z =0.2$ for
better rendering. Integration of these rates over $z$ and a 3--year
{\it LISA} mission lifetime yields a total of $N_{\rm gold} \sim 5$
golden events for the scenario with abundant MBHs and $N_{\rm gold}
\sim 1$ for the scenario with rare MBHs.
\label{fig:six}}
\end{figure}

\end{document}